\documentclass[pra,twocolumn,showpacs,preprintnumbers,amsmath,amssymb,revtex4-1]{revtex4-1}
\usepackage{amsfonts, amsmath, bm, bbm, graphicx, epsfig, epstopdf}
\usepackage{dcolumn}
\usepackage{textcomp}
 \usepackage[caption=false]{subfig}
 \usepackage[caption=false]{subfig}

    \usepackage{braket}
\renewcommand\bra[1]{{\langle{#1}|}}
\makeatletter
\renewcommand\ket[1]{%
  \@ifnextchar\bra{\k@t{#1}\!}{\k@t{#1}}%
}
\newcommand\k@t[1]{{|{#1}\rangle}}
\makeatother

\makeatother

\begin{document}
\title{ Universal terminal for mobile edge-quantum computing}

\author{Mohammadsadegh Khazali}
\affiliation{Institute for Quantum Optics and Quantum Information of the Austrian Academy of Sciences, A-6020 Innsbruck, Austria}
\affiliation{School of Physics, Institute for Research in Fundamental Sciences (IPM), Tehran 19395-5531, Iran}

\date{\today}

\begin{abstract}
To bring the quantum computing capacities to the personal edge devices,  the optimum approach is to have simple non-error corrected personal devices that offload the computational tasks to scalable quantum computers via edge servers with cryogenic components and fault-tolerant schemes. Hence the network elements deploy different encoding protocols. This article proposes quantum terminals that are compatible with different encoding protocols; paving the way for realizing mobile edge-quantum computing. By accommodating the atomic lattice processor inside a cavity, the entangling mechanism is provided by the Rydberg-Fermi cavity-QED technology. The gate operates by the Fermi scattering of a Rydberg electron from the plaquette atoms hosting the physical qubits. Therefore, different arrangements of logical-qubits derive the central atom over distinguished eigenstates, featuring photon emission at the early or late times distinguished by quantum interference. Applying an entanglement-swapping gate on two emitted photons would make the far-separated qubits entangled regardless of their encoding protocols. This gate provides a universal photonic interface for clustering the processors and connecting them with the quantum memories and quantum cloud that is compatible with different encoding formats.
\end{abstract}

\maketitle
\section{introduction}

Complicated quantum algorithms require large-scale fault-tolerant quantum processors. 
With the current noisy devices, the errors scale with the qubit numbers  to the limit that fails the operation of complicated tasks. 
The correction capabilities come with encoding the logical qubits in multiple physical qubits and are protected by error-correction codes \cite{Got97,Got99}. 
Quantum operations on the logic level are performed by an overload of operations at the level of physical qubits and require costly techniques such as complex optimization \cite{Her17,Han18}, magic state distillation \cite{Bra05}, transversal gates \cite{Got97} and lattice surgery \cite{Pou17,Gut19,Hor12}. 
This would dramatically increase the required computational power and capacity of the quantum processors. Considering the laboratory equipment for laser cooling and cryogenic environment, fault-tolerant quantum computation is not accessible on personal devices. 
To bring the quantum advantage to society, the optimized cost-benefit approach is to have  small-scale non-error corrected mobile devices that  offload the  computational tasks to the error-corrected servers \cite{Xu22}.
Hence the elements of the quantum cloud deploy different encoding protocols. A plug-\&-play approach for clustering and connecting these devices requires universal terminals that are compatible with all encoding schemes.

This paper proposes a gate that entangles a single photonic qubit with a logical-qubit encoded in 4-atoms \ \cite{Mar21} 
\begin{equation}
\label{Eq_Logical}
 \ket{0}_L=\frac{\ket{0000}+\ket{1111}}{\sqrt{2}}, \quad \, \ket{1}_L=\frac{\ket{0101}+\ket{1010}}{\sqrt{2}}.
 \end{equation}
The  extension to other encoding protocols e.g. 6-qubit  \cite{Sha08} is discussed.   
Applying a projective Bell-state measurement (PBM) on two flying qubits makes the qubits in stationary units entangled regardless of their encoding scheme, see Supp \cite{Supp}. 
This universal photonic interface is valuable for clustering the  fault-tolerant processors \cite{Mon14}, connecting them with the quantum memories \cite{Hes16,Kav13,Mon14}, and the quantum cloud \cite{Xu22,Jia22,Lok22,Zha22,LokSen22,Kris22,Ngu22,Nguy22,Khat22,Roh22,cac20,calef20,Chehimi22,Meter22,Dah22}.

\begin{figure}[h]
\centering 
\scalebox{0.55}{\includegraphics{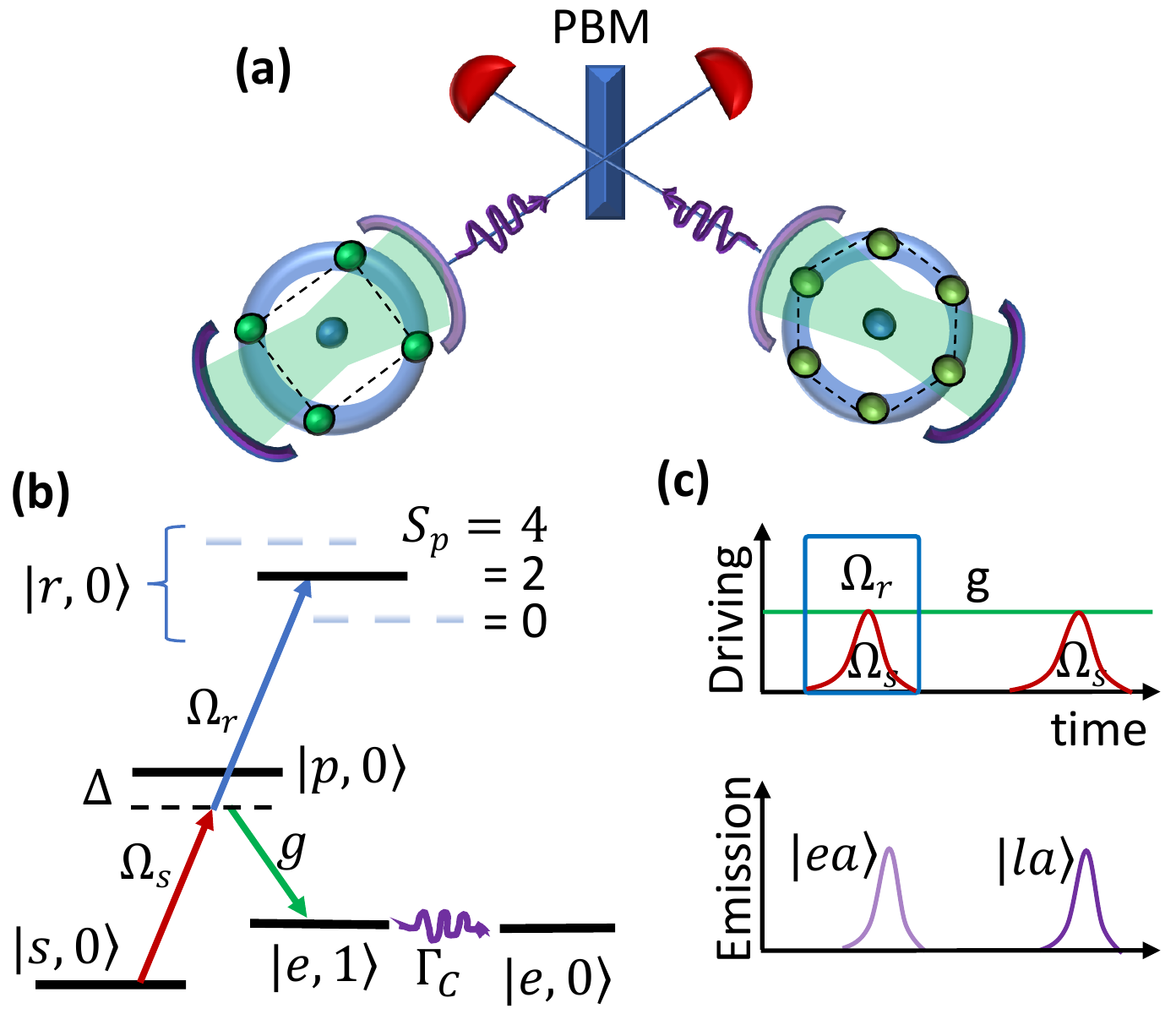}} 
\caption{Universal terminal for long-distance entanglement of logical qubits with different encoding formats. (a)  The auxiliary atom at the  center of the plaquette is emitting a photon at the early or late time, conditioned on the Rydberg-Fermi sensing of the logical qubit encoded  on plaquette atoms, see Fig.~2,4. Subsequent PBM of photons, entangles far-separated logical qubits \cite{Supp}.  (b) Level scheme of the auxiliary atom in the 4-qubit encoding scheme used in the left cavity (see Fig.~4 for the right cavity).  The lambda configuration consists of atom-laser and atom-cavity couplings and is responsible for photon emission. The interaction-induced level-shift of the Rydberg state depends on the plaquette's spin-state $S_p$. 
The $|0_L\rangle$ logical state is associated with $S_p=0$ and 4, making $\Omega_r$ laser out of resonance with the Rydberg level. Hence, the Raman transition from $|s\rangle$ to $|e\rangle$ would generate a single photon at the $|\text{early}\rangle$  time. On the other hand, $|1_L\rangle$ logical state accommodates two plaquette atoms in Rydberg wave-function $S_p=2$, making the $\Omega_r$ laser in-resonance with the Rydberg level. Consequently, destructive interference  inhibits the transition to $|e\rangle$ and therefore blocks the early photon emission. 
The following exclusive $\Omega_s$ pulse shown in  (c) leads to $|\text{late}\rangle$ photon emission in case of $|1_L\rangle$ state.} \label{Fig_Scheme}
\end{figure}

A logical qubit is a highly entangled two-dimensional subspace in the larger Hilbert space of multiple physical qubits. To avoid the  costly techniques used for logical operations \cite{Her17,Han18, Bra05,Got97,Pou17,Gut19,Hor12},  system-specific properties could be deployed to significantly reduce the number of operations on the physical qubits and hence the errors incurred during execution.   
Laser excited Rydberg atoms are an ideal example where the long-range interactions provide the possibility of simultaneous operations on multiple qubits \cite{Saf10,Kha20,Ise11,Gla17,Dla21,Kha16,Kha18,khaz2020rydberg}, with the bonus opportunities in quantum optics \cite{Ada19,Kha19,KhaRev21,Kha22}.
However, fast multi-qubit operation requires Rydberg population in all interacting qubits. This would affect the Rydberg stabilizer operations \cite{Gla17}  due to unwanted cross-talk between physical qubits with demolishing effects on the logical encoding.

The proposed gate scheme is based on cavity-QED photon emission from an auxiliary centered atom, conditioned on the logical-qubit that is encoded on the surrounding plaquette atoms. The single-step logical operation is performed by Fermi scattering of the central atom's Rydberg electron from the  plaquette atoms \cite{Gre00,Gaj14,Khaz21Fermi} in a spin-dependent lattice \cite{Dut98,Sol11,Kar09,Lee07,Jak99,Bre99,Bri00,Man03,Mandel03,Kum18}. Therefore, the logical-qubit determines which eigenstate the system would follow, either containing or excluding the atom-cavity coupling. 
As a result, the  logical-qubit would get entangled with the time-bin photonic qubit emitted by the central atom, generating the entangled state
\begin{equation}
\label{Eq_Gate}
(\ket{\text{early}}\ket{\text{0}_L}+\ket{\text{late}}\ket{\text{1}_L})/\sqrt{2}.
\end{equation}
 Subsequent PBM on two photons in time-bin basis \cite{Rie05, Mar03,Hal07,Tak09,Sun17,Boa18,Jin19}, entangles the  stationary-qubits regardless of their encoding schemes. 
 The time-bin qubit is an ideal choice due to intrinsic robustness against phase fluctuations \cite{Rie05}, with new advances in high-fidelity operations \cite{Bow21}.
 
 Deploying system-specific properties, the current scheme operates on all physical qubits in a single step,  circumventing the  costly techniques used for logical operations \cite{Her17,Han18, Bra05,Got97,Pou17,Gut19,Hor12}.
Remarkably,  the current proposal operates via a sole auxiliary Rydberg atom, eliminating the unwanted cross-talk of physical qubits over the logical operation.  Furthermore, the Rydberg-Fermi interaction provides a molecular type potential that eliminates the frozen gas regime requirement of the Rydberg dipolar schemes and also operates at much shorter interatomic distances, addressing the scalability. Localizing the Rydberg population on one atom in this scheme, closes the collective line broadening problems \cite{Gol16,Zei16}. Unlike the majority of Rydberg dipolar schemes that operates via a $\pi$-gap-$\pi$ scheme, the proposed Ryd-Fermi gate does not leave any Rydberg population unprotected from the laser, closing the major decoherence source \cite{Mal15,Gra19}. Suppressing the short-lived Rydberg population, high-fidelities of 99.6\% is predicted.

\section{scheme}

The setup consists of cesium $^{133}$Cs  atomic lattice quantum processor accommodated in a cavity.
Fig.~\ref{Fig_Scheme}a, is a simplified figure with a single plaquette and a  central atom in two edge devices of the network. The logical information in two devices is encoded in different 4- and 6-qubit basis. The system atoms are placed on square (hexagonal) plaquette consisting of physical qubit states $\ket{0}=\ket{6S,F=3}$ and $\ket{1}=\ket{6S,F=4}$, while auxiliary atom responsible for conditional photon emission is at the center of the plaquettes with the electronic level scheme presented in  Fig.~\ref{Fig_Scheme}b (and Fig.~4b in case of 6-qubit encoding). 
The lower lambda configuration is responsible for on-demand single-photon emission, similar to \cite{Kuh99,Gog10}. 
The $\Lambda$ transition would be  (allowed) prohibited due to the quantum interference in case the blue laser is  (off-resonance) in-resonance with the Rydberg transition. 
The tuning of the blue laser is determined by the Fermi scattering of the Rydberg electron from the plaquette atoms, designed to make the photon emission conditioned on the logical qubit. 
In the lower lambda system the transition between $\ket{s}=\ket{6S_{1/2}, F=3}$ and $\ket{p}=\ket{7P_{3/2}}$ states is derived by $\Omega_s$ laser while the $\ket{p}$ to $\ket{e}=\{\ket{6S_{1/2}, F=4} \, \text{or}\, \ket{5D_{3/2}}\}$ is governed by the Jaynes-Cummings interaction with the coupling constant $g$ and the cavity mode with frequency  $\omega_c\approx\{450,\, \text{or}\,1300\}$nm \cite{Microwave}.  The transitions are detuned from the intermediate state by $\Delta$. The state basis $\ket{i,n}$ in Fig.~\ref{Fig_Scheme}b includes the central atom's electronic state $\ket{i}$ and cavity number state $\ket{n}$. 
Over a Raman $\pi$ pulse, the transition from $\ket{s,0}$ to $\ket{e,1}$ would generate a photon in the cavity. The decay of the cavity mode causes single-photon emission and the system settles itself in the state $\ket{e,0}$.

Fermi scattering of central atom's Rydberg electron from plaquette atoms in spin-dependent lattice results in an effective level-shift quantified by the plaquette spin $S_p=\sum_{j\in p} \sigma_{00}^{(j)}$, with $\sigma_{00}=\ket{0}\bra{0}$ being the projective operator and $j$ goes over the plaquette atoms, see  Fig.~\ref{Fig_Scheme}b. Having logical qubit $\ket{0}_L=(\ket{0000}+\ket{1111})/\sqrt{2}$, the blue laser would be out of resonance with the interaction-shifted Rydberg level. Hence, the first  $\Omega_s$ pulse in Fig.~1c derives an early photon emission $\ket{\text{early}}$.
For the other logical qubit $\ket{1}_L=(\ket{0101}+\ket{1010})/\sqrt{2}$, two plaquette atoms would be in Rydberg electron wave-function. Consequently, the blue laser would get in resonance with the Rydberg level, blocking the photon emission over the first  $\Omega_s$ pulse.  The remaining $\ket{s,0}$ population would then get transferred  over the second  $\Omega_s$ pulse (not accompanied by the Rydberg lasers $\Omega_r$)  and hence emits a late photon $\ket{\text{late}}$, leading to the desired state of Eq.~\ref{Eq_Gate}. The time ordering of the pulses is shown in Fig.~\ref{Fig_Scheme}c, and the physics behind the blocking and transmission are discussed below.
In an entanglement swapping station, the emitted photons from distinct sources would undergo  a C-NOT gate followed by projective Bell state measurement. This would make the two far-separated plaquettes entangled in their own logical basis \cite{Supp}.

\begin{figure}
\centering 
       \scalebox{0.37}{\includegraphics{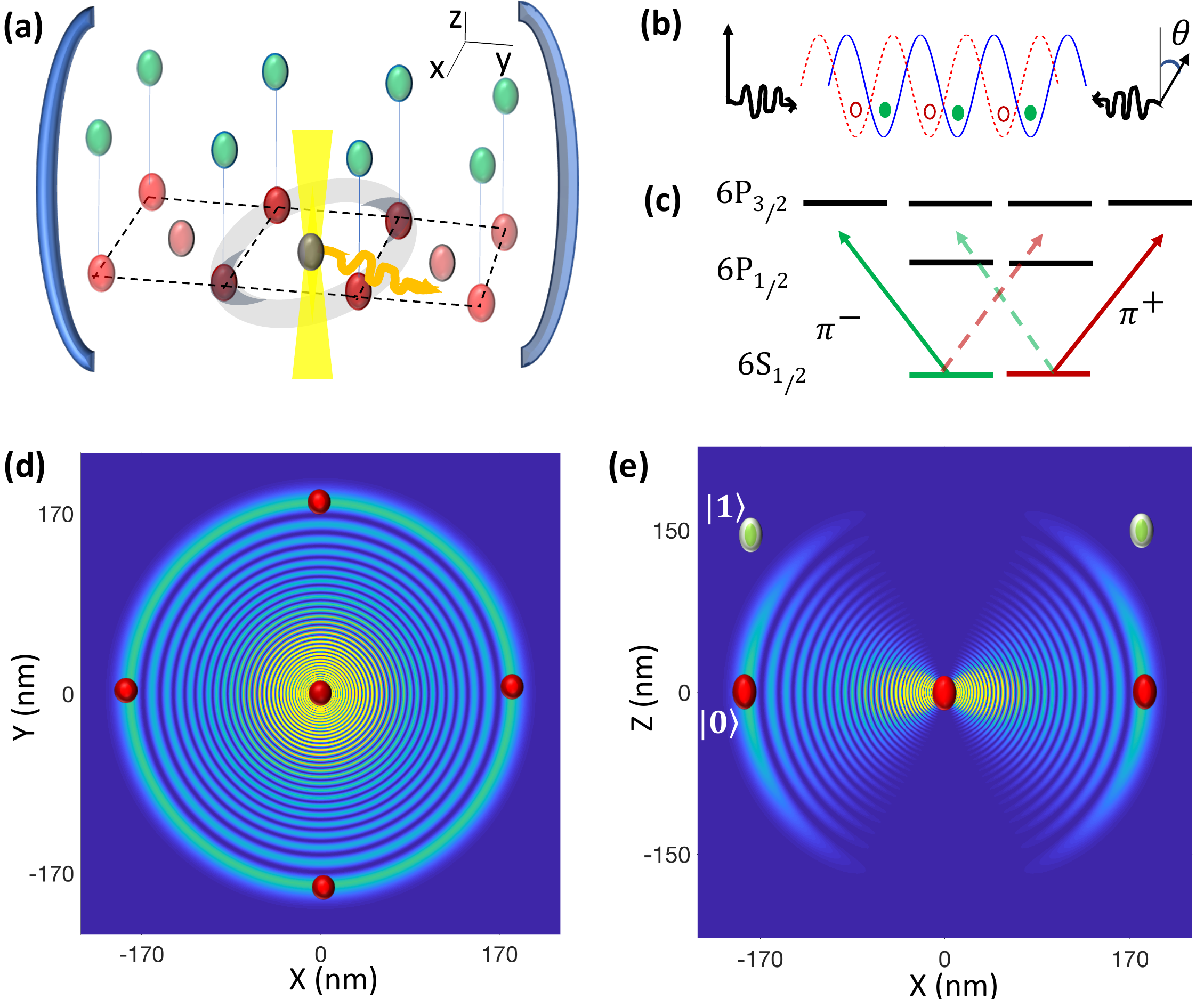}} 
        \caption{Rydberg-Fermi interaction in spin-dependent lattice. (a) In a 2D lattice with a single atom per site, applying a qubit-dependent lattice-shift along $z$ makes each atom in a spatial superposition of being in red and green sites where the components are controlled by the internal electronic qubit-states $|0\rangle$ and $|1\rangle$. (b) Counter propagating linearly polarized lights along $z$ direction with relative polarization shift of $\theta$  forms two standing-waves of  $\pi^{-}$ and $\pi^{+}$ circular polarizations. (c) Tunning the trapping laser between  $6P_{3/2}$ and $6P_{1/2}$, the polarizability of qubit states $|0\rangle$ and $|1\rangle$ are given by distinguished circularly polarized lights $\pi^{-}$ and $\pi^{+}$ respectively, resulting to spin-dependent trapping. 
The  cross-sections of $|45D_{5/2},5/2\rangle$ Rydberg wave-function along (d) xy and (e) xz are plotted. The position of the qubit states $|0\rangle$ and $|1\rangle$ are marked by the red and green circles.  (e) Note to the spin-dependent overlap of plaquette atoms with the  Rydberg electron wave-function. 
 }\label{Fig_RydFermi}
\end{figure}

\section{Rydberg-Fermi interaction}
\label{Sec_RydFermi}
 To apply a qubit-dependent Rydberg-fermi interaction, the lattice undergoes spin-dependent shift along the $z$ direction. 
Hence, the Rydberg electron of the central atom would only scatter from the plaquette atoms in qubit state $\ket{0}$. This would cause a level-shift that depends on the plaquette spin   $S_p=\sum_{j\in p} \sigma_{00}^{(j)}$.

The {\it spin-dependent lattice} shift along $z$ direction is formed by counter-propagating linearly polarized lights as depicted in Fig.~\ref{Fig_RydFermi}b.
Having a relative shift between the fields' polarizations $2\theta$, the total electric field could be written in terms of the sum of right  and left circularly polarized lights $E=E_0 \exp(-i \nu t)(\pi^{+}\sin(kz+\theta)+\pi^{-}\sin(kz-\theta))$. 
To make a spin-dependent lattice-shift, the spin polarizabilities should be linked to different circular polarization components of lights \cite{Dut98}. To cancel the polarizabilities with unwanted light elements shown by dashed lines in Fig.~\ref{Fig_RydFermi}c, the  trapping laser must be tuned between $P_{3/2}$ and $P_{1/2}$ states so that the ac-Stark shifts of these two levels cancel each other.
As a result the $m_j=\pm1/2$ levels of the ground state would be trapped by $V_{\pm}=\alpha |E_0|^2 \sin(kz\pm\theta)$ respectively.
The hyperfine qubit states $|0\rangle=|F=3,m_F=3\rangle$ and  $|1\rangle=|F=4,m_F=4\rangle$ would experience $V_{\ket{0}}=(V_{+}+3V_{-})/4$, $V_{\ket{1}}=V_{\ket{s}}=V_{+}$.

The Rydberg laser $\Omega_r$ is exciting the auxiliary atom at the centere of the desired plaquette, see Fig.~\ref{Fig_RydFermi}. Applying a spin-dependent lattice shift perpendicular to the 2D lattice would result in dual spin/spatial encoding of the plaquette qubits. 
Thus, exciting the central atom to Rydberg level, the electron wave-function would   exclusively overlap with plaquette qubits in spin-state $\ket{0}$, see Fig.~\ref{Fig_RydFermi}e. The resulting {\it Rydberg-Fermi interaction} caused by scattering of Rydberg electron from a single neutral atom would be quantified by  \cite{Fer,Eil17,Eil19},
\begin{equation}
\label{Eq_RydFermi}
V_{\text{RF}}=(2\pi \frac{\tan(\delta^s)}{k(R)}-6\pi\frac{\tan(\delta^p)}{k^3(R)}\stackrel{\leftarrow}{\nabla}_{{\bf r}}.\stackrel{\rightarrow}{\nabla}_{{\bf r}})\delta({\bf r}-{\bf R})
\end{equation}
with {\bf r} and {\bf R} being the positions of the Rydberg electron and a ground state atom with respect to the ionic core, and $\delta^{\{s,p\}}$  are the triplet s- and p-wave scattering phase shift of Rydberg electron from a neighboring ground state atom \cite{khu02}. Since $S_p$ presents the number of plaquette atoms in the Rydberg wave-function of central atom, the effective level-shift caused by the Fermi scattering would be $S_pV_{\text{RF}}$ \cite{Gaj14}, see Fig.~\ref{Fig_Scheme}b.

\section{entangling gate}
\label{Sec_PhotPar}

\begin{figure}
\centering 
       \scalebox{0.315}{\includegraphics{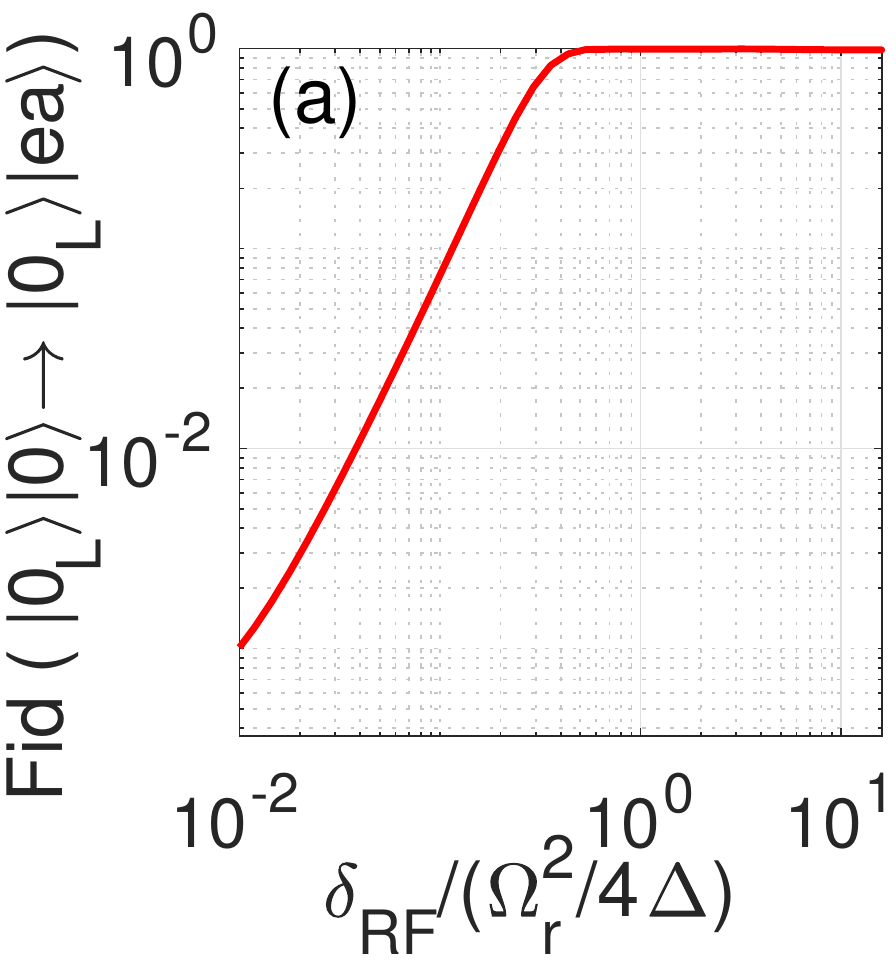}}        
       \scalebox{0.315}{\includegraphics{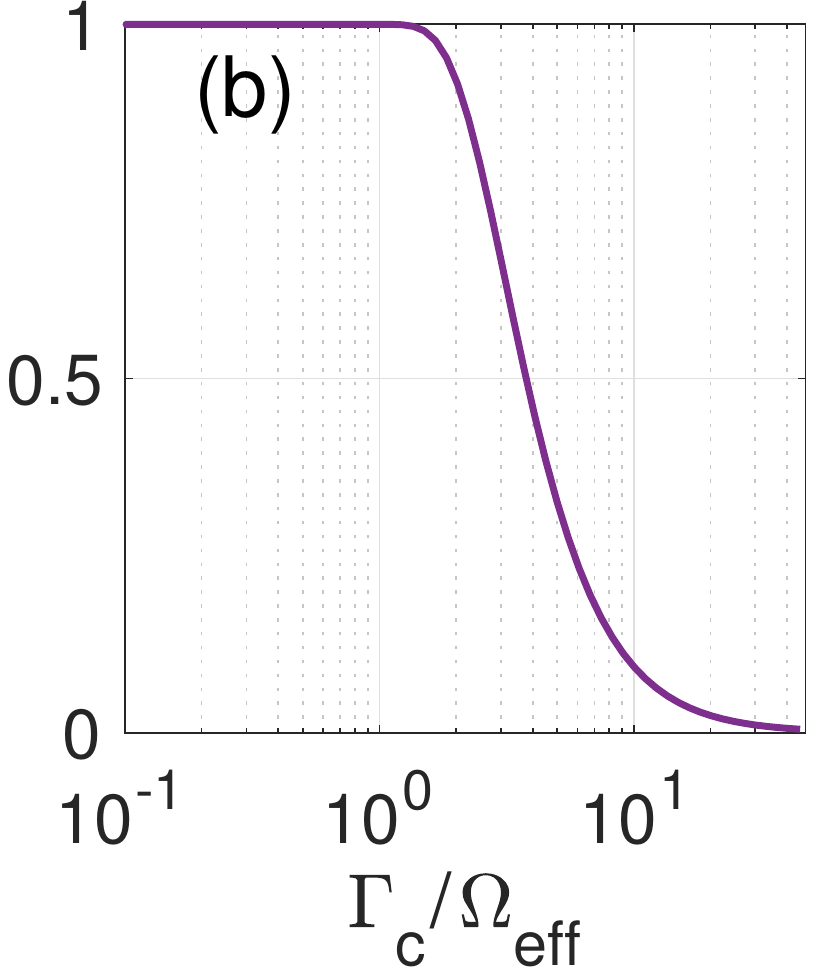}} 
              \scalebox{.32}{\includegraphics{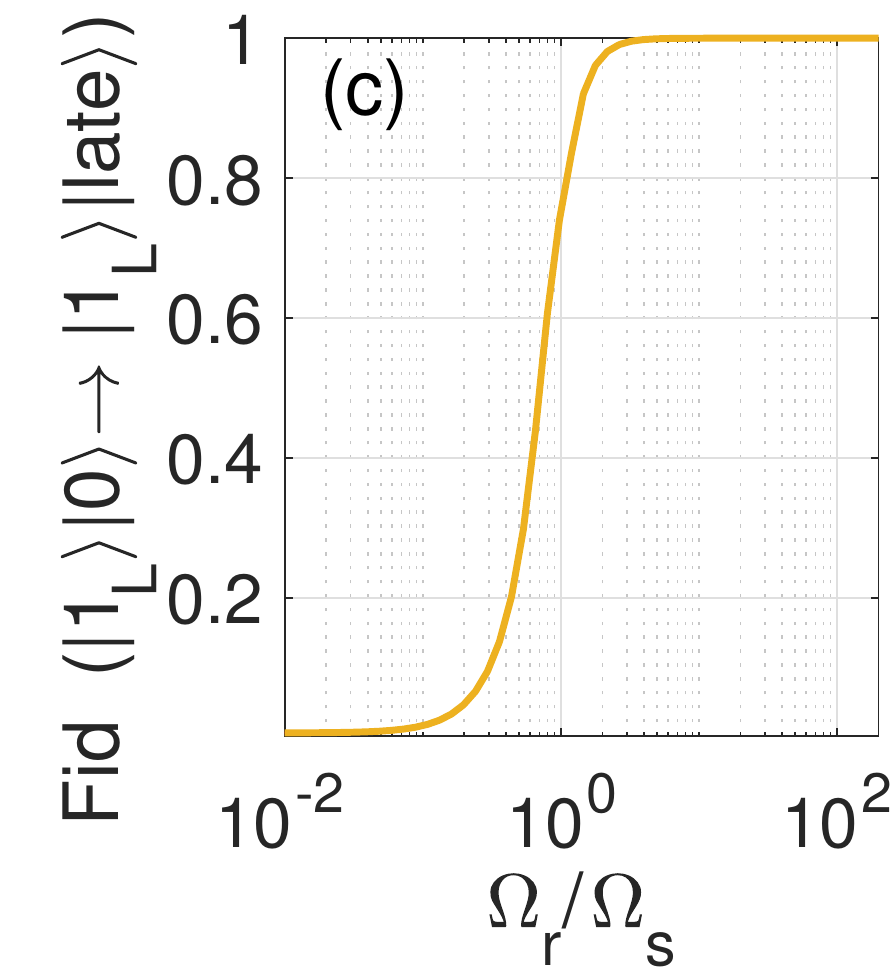}}        
        \caption{The operation regime of entangling gate. The operation is simulated under the non-hermitian Hamiltonian $\tilde{H}$ of Eq.~\ref{Eq_Htilde}. (a) With  $|0_L\rangle$, the detuning $\delta\gtrsim 1$ caused by Ryd-Fermi interaction lifts the interference of dark and bright states, allowing early photon emission.
        (b) With $|0_L\rangle$ logical state ($\delta_{RF}=2V_{RF}$),  the optimum early emission occurs at  $\Gamma_c=\Omega_{\text{eff}}$. Larger cavity leakage would reduce the coherence, scatter the population out of the dark state, and reduce the emission probability.  (c) Having $|1_L\rangle$ logical state ($\delta_{RF}=0$), the early photon emission would be blocked for $\Omega_r/\Omega_s>3$ by more than 99\%. 
Applied parameters are $\delta_{RF}/2\pi=130$MHz, $\Delta/2\pi=33$MHz, $\Omega_r/2\pi=40$MHz, $\max(\Omega_s)/2\pi=10$MHz, $g/2\pi=8.5$MHz, $\Gamma=\Omega_{\text{eff}}$, $\sigma\Gamma=10$.
}\label{Fig_PhNum}\end{figure}

This section discusses the main physics behind the { conditional photon emission} that entangles the stationary logical and flying qubits. 
In the case of $\ket{1_L}$, the physical qubits' arrangement in the qubit-dependent lattice would accommodate two plaquette atoms in the Rydberg wave-function of the central atom,  making the    $\Omega_r$ laser in-resonance with the shifted Rydberg level $\delta_{\text{RF}}(\ket{1_L})=0$. 
Having $\ket{0_L}$, the plaquette's arrangement contains either $S_p=0$ or 4 atoms in the  Rydberg wave-function, resulting in the effective detuning of $\delta_{\text{RF}}(\ket{0_L})=2V_{\text{RF}}$.
The  Hamiltonian of the central atom in the electronic basis is given by
\begin{eqnarray} 
 \label{Eq_H}
H=&&\Omega_s/2(\hat{\sigma}_{sp}+\text{h.c.}) + \Omega_R/2(\hat{\sigma}_{rp}+\text{h.c.}) \\ \nonumber 
&&+ g/2(\hat{\sigma}_{ep}+\text{h.c.})+{\Delta}\hat{\sigma}_{pp}+\delta_{RF}\hat{\sigma}_{rr}  \quad\quad\quad
\end{eqnarray}  
 where $\hat{\sigma}_{\alpha \beta}=\ket{\alpha}\bra{\beta}$.
 To simplify the analytic discussion here the large detuning regime $\Delta\gg \Omega_{\{s,r\}},g$ and equal couplings $g=\Omega_s$ are considered. 
 After adiabatic elimination of the intermediate state $\ket{p}$,  the Hamiltonian is represented in new basis $\ket{\pm}=(\ket{s}\pm\ket{e})/\sqrt{2}$, and $\ket{r}$ as
 \begin{eqnarray} 
 \label{Eq_HNoP}
H/\epsilon=\lambda^2 \ket{+}\bra{+} + (1-\delta) \ket{r}\bra{r} + \lambda (\ket{+}\bra{r}+\text{h.c.})\quad \quad  \quad  
\end{eqnarray}  
where $\lambda=\Omega_s/\Omega_r$ is the dimensionless Rabi frequency and $\delta=\delta_{RF}/\epsilon$ is the Rydberg-Fermi interaction scaled by $\epsilon=\Omega_r^2/4\Delta$.  
 In the regime of $\lambda,\delta\ll1$ the lower  bright $\ket{b}=[(1-\delta)\ket{+}-\lambda\ket{r}]/\sqrt{(1-\delta)^2+\lambda^2}$ state, with the energy $2\delta \lambda^2$ would interfere with the dark state $\ket{d}=\ket{-}$. 
 Having  $\ket{1_L}$, the Rydberg laser would be in resonance $\delta=0$. Hence, the system would follow the new dark state $\ket{D}=(\ket{d}+\ket{b})/\sqrt{2}=(\ket{s}-\lambda\ket{r})/\sqrt{1+\lambda^2}$ featuring destructive interference that blocks the transition to $\ket{e}$ state.
A Ryd-Fermi induced level-shift  of $\delta=\delta_{\text{RF}}/(\Omega_r^2/4\Delta)\gtrsim 1$  lifts the $\ket{d}$ and $\ket{b}$ states degeneracy, the system would exclusively follow the dark state $\ket{d}$ and emits an early photon, see Fig.~\ref{Fig_PhNum}a.

While a large cavity decay rate $\Gamma_c$ enhances the emission rate and narrows the emission probability profile, it  reduces the coherence. 
In the regime of $\lambda=\Omega_s/\Omega_r\ll1$   the steady-state of the master equation could be obtained perturbatively as $\rho=\rho_0+\lambda \rho_1+\lambda^2 \rho_2$. 
The desired transition coherence is given by 
\begin{equation}
\rho_{se}= \frac{\Omega_s g}{4\Delta} \frac{2i \delta_{RF}}{\Gamma_c(\Omega_r^2/4\Delta - \delta_{RF})}.
\end{equation}
Cavity decay rates $\Gamma_c$ larger than effective transition Rabi frequency $\Omega_{\text{eff}}=\Omega_s g/4\Delta$  reduces the coherence between $\ket{s}$ and $\ket{e}$. Since an incoherent superposition of these states does not exclusively project into the dark state $\ket{d}$, some of the population would transfer to bright states reducing the operation fidelity, see Fig.~\ref{Fig_PhNum}b. Finally, the $\lambda=\Omega_s/\Omega_r\ll1$ condition is required to block the early transmission of photons in the presence of $\ket{1_L}$  as numerically evaluated in Fig.~\ref{Fig_PhNum}c.

To study the {\it Photon emission},  Eq.~\ref{Eq_H} must be modified to the non-hermitian version
\begin{equation}
\label{Eq_Htilde}
\tilde{H}=H-\text{i}\Gamma_c\hat{\sigma}_{ee},
\end{equation}
where $\Gamma_c$ is the cavity leakage rate.
The gate operation under this Hamiltonian is quantified in Fig.~\ref{Fig_PhNum}.
In the regime that $\Delta,\delta_{RF}\gg\Omega_{\{r,s\}},g$, the adiabatic elimination of Rydberg and intermediate $p$ state would simplify the dynamics to a two-level system with an effective Rabi frequency of $\Omega_{\text{eff}}=\frac{\Omega_s g}{4{\Delta}} (1+\frac{\Omega_r^2}{4{\Delta}\delta_{RF}})$ and an effective detuning of $(\frac{\Omega_s^2}{4\Delta}-\frac{g^2}{4\Delta})(1+\frac{\Omega_r^2}{4{\Delta}\delta_{RF}})+\text{i}\Gamma_c/2$. Choosing close coupling strengths $\Omega_s\approx g$ to facilitate the desired transition,  the photon flux out of the cavity would be given by $\Phi(t)=\Gamma_c|c_{e,1}|^2=\Gamma_c\frac{\Omega_{\text{eff}}^2 }{\tilde{\Omega} ^2}  e^{-\Gamma_c t} \sin^2(\tilde{\Omega} t)$ where $\tilde{\Omega}=\sqrt{\Omega_{\text{eff}}^2+\Gamma_c^2/4}$.

\section{Implementation}

The {\it implementation} of the scheme encounters a 340nm in-plane $xy$  optical lattice addressing $6S-10P$ transition with trap potential of $U_{\text{trap-xy}}/2\pi=5$MHz.  The spin-dependent trapping along $z$ direction is formed by 870nm laser with $U_{\text{trap-z}}/2\pi=2$MHz, see Fig.~\ref{Fig_RydFermi}c. 
Targeting the central atom  to  $|45D_{5/2},5/2\rangle$ and considering the spin-dependent lattice shift of  $D_z=150$nm, the average Rydberg-Fermi interaction with a single plaquette atom in qubit-state $\ket{0}$ would be $\bra{0}V_{RF}\ket{0}=2\pi\times 65$MHz \cite{Supp}. The atoms in the other qubit-dependent lattice experience negligible interaction of $\bra{1}V_{RF}\ket{1}=2\pi\times 1$kHz.  
The optimum realistic parameters in driving of the central atom includes $\Omega_r/2\pi= 40$MHz, max$(\Omega_s)/2\pi=10$MHz, $g/2\pi= 8.5$MHz, $\Gamma_c/2\pi=750$kHz \cite{Kuh99}, $\Delta/2\pi=33$MHz and $\delta_{\text{RF}}(\ket{0_L})/2\pi=130$MHz. 
The  time interval of a Gaussian pulse $\Omega_s(t)=\Omega_s \exp(-t^2/\sigma^2)$ must be long enough  1- to fulfill the adiabaticity condition and preserve the dark state $\sigma \gg g^{-1}$; 2- to ensure the photon emission considering the cavity decay-time   $\sigma \gg \Gamma^{-1}$;  and 3- to perform the population rotation  $\sigma\gg \Omega_{\text{eff}}^{-1}$. Considering the spontaneous emission of the intermediate $\gamma_{p}/2\pi=1$MHz \cite{Toh19} and the Rydberg state  $\gamma_{r}/2\pi=4.5$kHz \cite{Bet09} and $\sigma=3.5\mu$s, the operation fidelity of $\ket{0}(\ket{{0_L}}+\ket{{1_L}})/\sqrt{2}\rightarrow(\ket{\text{early}}\ket{0_L}+\ket{\text{late}}\ket{1_L})/\sqrt{2}$ for the mentioned set of parameters under the numerical simulation would be Fid=99.6\%. 
Further nuances regarding single-site addressing, perseverance of ground motional state, and Rydberg-molecule decoherence channel are discussed in Supp \cite{Supp}.

\section{extension to six-qubit encoding}

\begin{figure}
\centering 
\scalebox{0.5}{\includegraphics{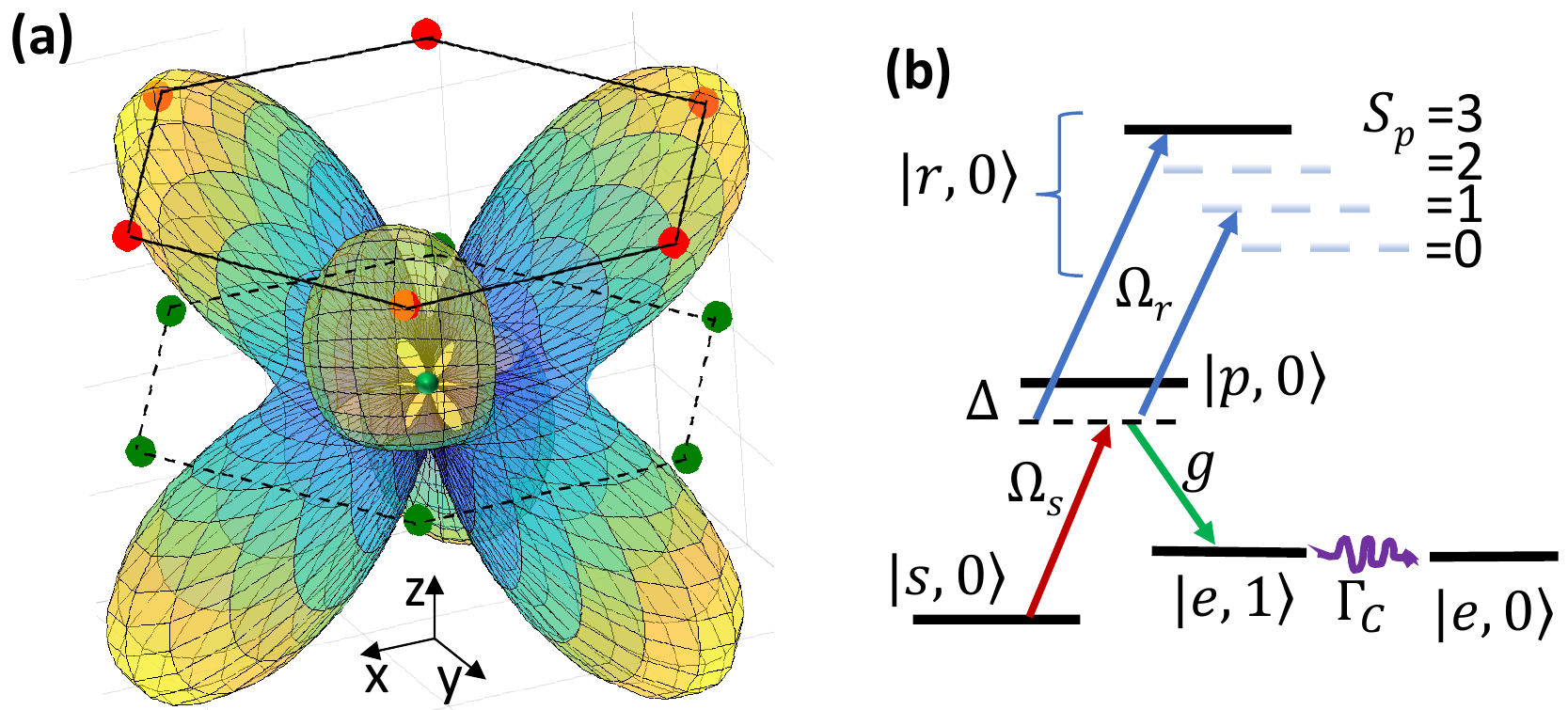}} 
\caption{Photonics interface for 6-qubit surface code. The physical qubits are dual encoded in the spin/spatial basis of a spin-dependent lattice placed over the hexagonal structure. (a) The logical operator $\bar{Z}=I Z I Z I Z$ is implemented by exciting the auxiliary atom at the centre of hexagonal plaquette to the Rydberg superposition state  with angular part $(Y_2^2(\theta,\phi)+Y_{2}^{-1}(\theta,\phi))/\sqrt{2}$.  (b) The energy splitting of the Rydberg state depends on the number of atoms in the Rydberg orbitals  $S_p=\sum_{i=2,4,6}\sigma_{00}^{(i)}$. The two-colored Rydberg lasers $\Omega_r$ get in resonance with the odd parity of $\bar{Z}$, associated with $|1\rangle_L$, blocking the early photon emission.  } \label{Fig_6Qubt}
\end{figure}

The current proposal could be extended to the {\it six-qubit code}  \cite{Sha08}, with the  logical basis being the odd and even parity states of the  $\bar{Z}=I Z I Z I Z$ logical-operator \cite{qubit}. 
Exciting Rydberg superposition state  $R_{n,l}(r)(Y_2^2(\theta,\phi)+Y_{2}^{-1}(\theta,\phi))/\sqrt{2}$, could realize the desired  parity-photon entangling gate in a triangular lattice, see Fig.~\ref{Fig_6Qubt}a.  
Fermi scattering of central atom's Rydberg electron from every other site in a hexagonal plaquette structure is conditioned on the physical qubits to be in $\ket{0}$ state. This would cause an effective level-shift quantified by three atoms' spin-number, see Fig.~\ref{Fig_6Qubt}b. 
The two-color  $\Omega_r$ in Fig.~\ref{Fig_6Qubt}b are tuned in-resonance with odd-parity of the  $\bar{Z}$ stabilizer. The two-color blue transitions could be obtained by a single laser in a setup of beamsplitters and acusto-optical modulators.
With even parity of $\bar{Z}$, the blue laser would be out of resonance with the Rydberg level. Hence, the first  $\Omega_s$ pulse in Fig.~1c derives an early photon emission $\ket{\text{early}}$.
Having an odd number of spin-downs in  the three targeted atoms, makes one of the blue lasers in resonance with the Rydberg level,
while the other laser causes a level-shift to $\ket {p}$ state resulting in a small correction in the detuning value  $\Delta\equiv\Delta \pm \Omega_r^2/4V_{\text{RF}}$. The resonant laser would block the transition over the first  $\Omega_s$ pulse as discussed in Sec.~\ref{Sec_PhotPar}. The remained $\ket{s,0}$ population would then transfer over the second  $\Omega_s$ pulse (not accompanied with $\Omega_r$)  and hence emits a late photon $\ket{\text{late}}$, providing the desired state of Eq.~\ref{Eq_Gate} in the 6-qubit encoding. 

\section{Conclusion and Outlook}

This article proposes a hardware architecture and protocol for long-distance entanglement generation between logical qubits with different encoding protocols. 
The Fermi-scattering of the Rydberg electron from the plaquette atoms in a spin-dependent lattice forms a level-shift that is quantified by the logical qubit. This level-shift would control the central atoms coupling with the cavity mode and hence the ordering of photon emission would be entangled by the logical qubit. The two characteristic of the Rydberg-fermi interaction in working at small interatomic distance and in making bound state with restoring force allows long-time operation in ultra-dense atomic processors, making a significant advantage over the Rydberg dipolar counterparts.
The proposed photonic quantum bus solution, promises the scalability of fault-tolerant processors at the current NISQ-era devices. This would resemble the integrated
circuits (IC) technology of silicon-based processors characterized by Moore's law \cite{Moo69}. 
With the transition/processing limits of   the quantum satellites \cite{Dir21,Yua21,Nit22,Min22} and mobile edge devices,  communication via a single photon is desired. With large  communication channel bandwidth, emitted photons from the gate could be encoded into three-photons GHZ state \cite{Ash85,Yos18} to correct the bit-flip, while the photon loss could be neglected by post-selection after the PBM. 
From a fundamental perspective, the presented technique of entangling multiple physical qubits in the stabilizer basis facilitates the investigation of phenomena, properties, and protocols that arise in quantum information over a wide dimension of state space, while operating on the basis that is growing polynomially.

\begin{widetext}

\section*{Supplemental document for ''Universal terminal for mobile edge-quantum computing''}

\section*{S1.Rydberg Fermi interaction}

This section discusses, the steps in calculating the potential energy curves (PEC) under the Rydberg-Fermi interaction \cite{Fer,Eil17,Eil19}
\begin{equation}
\label{Eq_RydFermi}
V_{\text{RF}}=(2\pi \frac{\tan(\delta^s)}{k(R)}-6\pi\frac{\tan(\delta^p)}{k^3(R)}\stackrel{\leftarrow}{\nabla}_{{\bf r}}.\stackrel{\rightarrow}{\nabla}_{{\bf r}})\delta({\bf r}-{\bf R}).
\end{equation}
 The coupling of the Rydberg state $|45D_{5/2},5/2\rangle$ with the neighbouring states  $|43H+45D_{3/2}+46P_{\{1/2,3/2\}}+47S_{1/2}\rangle$ is considered at the position of neighbouring lattice sites $R$. Here $\ket{nH}=\sum_{l,m} \ket{n,l,m}$ represents the Hydrogen state encountering semi degenerate orbital angular momentum numbers $2<l<n$.
 The matrix elements in the  manifold of coupled states are given by
\begin{eqnarray}
&&H_{nlm,n'l'm'}({\bf R})=\bra{\psi_{nlm}({\bf R})} V_{\text{RF}}\ket{\psi_{n'l'm'}({\bf R})} \quad \quad \quad \quad \\ \nonumber
&&H_{nlm,nlm}=-\frac{Ry}{n^{*2}}\\ \nonumber
\end{eqnarray}
where $Ry$ is the Rydberg constant of $Cs$ atoms and $n^{*}$ is the effective Rydberg principal number.
Diagonalizing 8000 coupled states, the energy potential is plotted in Fig.~\ref{Fig_PEC}. In the UV optical-lattice with $\lambda=350$nm, the Fermi scattering of Rydberg electron from the neighbouring lattice site would result in 400MHz level-shift of the Rydberg level ideal for fast quantum operations.

  \begin{figure}[h]
\centering         \scalebox{0.45}{\includegraphics{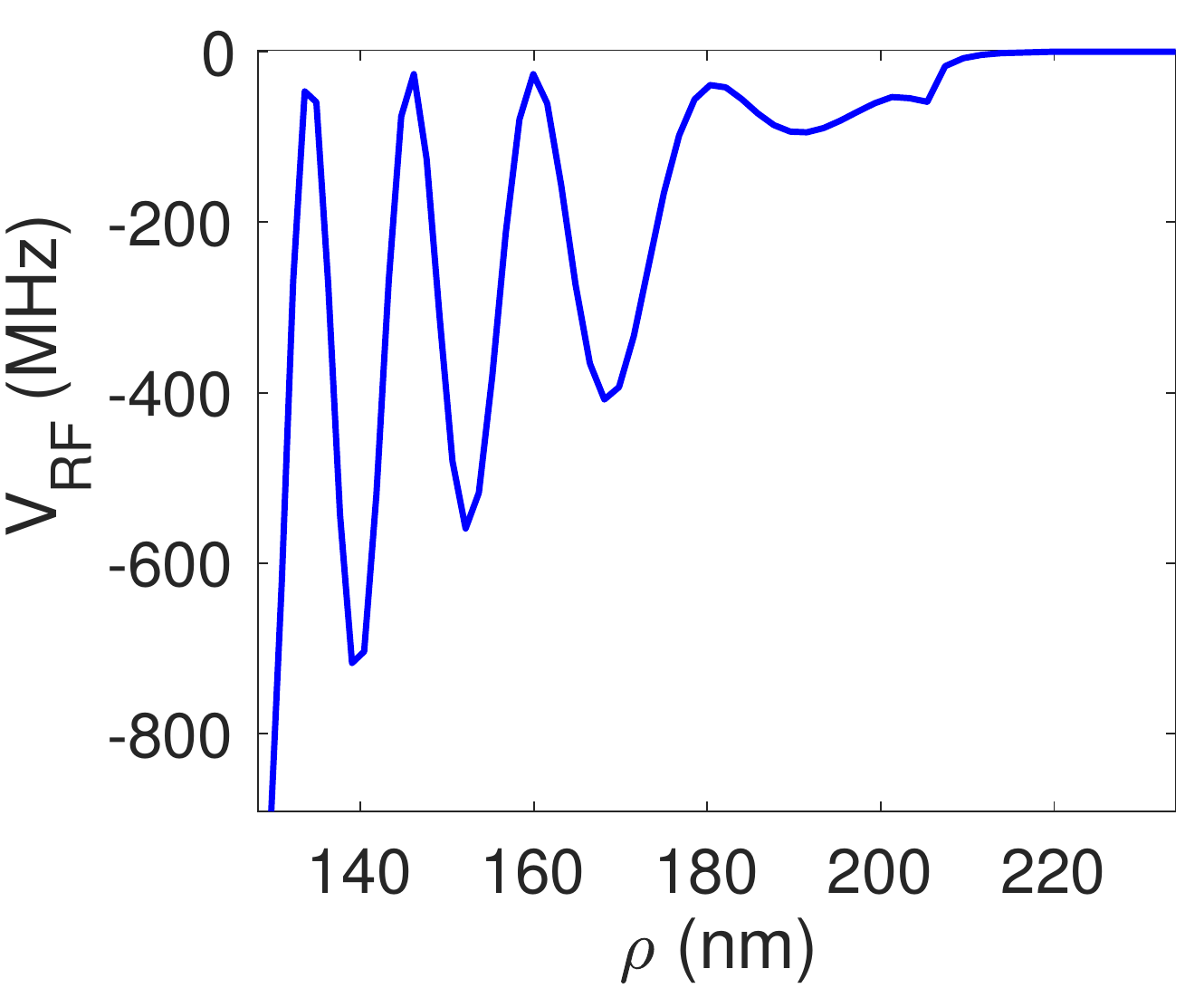}} 
\caption{ PEC with S- and P-wave scattering in Cs atoms. Here the coupling of the Rydberg state $|45D_{5/2},5/2\rangle$ with the neighbouring states  $|43H+45D_{3/2}+46P_{\{1/2,3/2\}}+47S_{1/2}\rangle$ is considered under Eq.~\ref{Eq_RydFermi}. Interaction strength is plotted across radial direction with $\theta=\pi/2$. 
}\label{Fig_PEC}
\end{figure}

For the p-wave scattering of Rydberg electron from the neighbouring ground state atom, the gradient of the Rydberg wave-function $\psi=R_{nl}(r)Y_l^m(\theta,\phi)$ at the position of the neighbouring lattice site is required which is
\begin{equation}
 \nabla \psi(r,\theta,\phi)=\begin{bmatrix} 
	\frac{\partial R_{nl}}{\partial r}Y_l^m\\
	\frac{1}{r}R_{nl}\frac{\partial Y_l^m}{\partial \theta} \\
	\frac{1}{r\sin\theta}R_{nl}\frac{\partial Y_l^m}{\partial \phi} \\
	\end{bmatrix}=\quad\quad\quad\quad\quad\,\quad~\,\,\,\,\,\,\,\,\,\,\,\,\,\,\,\,\,\,\,\,\,\,\,\,
	\end{equation}
\begin{equation}
\begin{bmatrix} \nonumber
	\frac{\partial R_{nl}(r)}{\partial r}Y_l^m(\theta,\phi)\\ \nonumber
	\frac{1}{r}R_{nl}(r)\frac{1}{2} \sqrt{l^2-m^2}[Y_l^{m+1}(\theta,\phi)e^{-\text{i}\phi}- (l+m+1)Y_l^{m-1}(\theta,\phi)e^{\text{i}\phi}] \\ \nonumber
	\text{i}m\frac{\psi(r,\theta,\phi)}{r\sin(\theta)}\\ \nonumber
\end{bmatrix} \nonumber
\end{equation}
in the spherical coordinate. The radial wave-function and its derivative are calculated numerically using Numerov technique \cite{Gal05}.

\section*{S2. Entanglement swapping}
\label{Sec_EntSwap}
Entangling far separated atomic qubits in the logical basis requires Bell state projective measurement of the cavity emitted photons.
The proposed Rydberg-Fermi cQED gate in this article would provide atom-photon entanglement in each cavity setup, see top line of Eq.~\ref{Eq_ent}.  The state could then rearranged in a separate logical and photonic qubit pairs in the following line
\begin{eqnarray}
\label{Eq_ent}
\ket{\psi}=\frac{(\ket{1_L}\ket{\text{late}}+\ket{0_L}\ket{\text{early}})_1}{\sqrt{2}}\frac{(\ket{1_L}\ket{\text{late}}+\ket{0_L}\ket{\text{early}})_2}{\sqrt{2}} \\ \nonumber 
=(|{\tilde{\phi}}^{+}_L\rangle |\tilde{\phi}^{+}_p\rangle+ |{\tilde{\phi}}^{-}_L\rangle |\tilde{\phi}^{-}_p\rangle +|{\tilde{\psi}}^{+}_L\rangle |\tilde{\psi}^{+}_p\rangle +|{\tilde{\psi}}^{-}_L\rangle |\tilde{\psi}^{-}_p\rangle )/2, 
\end{eqnarray}
where the rotated Bell states are obtained by applying a Hadamard on the second element i.e.  $|\tilde{\phi}^{\pm}\rangle = \ket{1}_1\ket{+}_2\pm \ket{0}_1\ket{-}_2$ and $|\tilde{\psi}^{\pm}\rangle= \ket{0}_1\ket{+}_2\pm \ket{1}_1\ket{-}_2$ where $|\pm\rangle =( \ket{1}\pm|0\rangle)/\sqrt{2}$.  The photonic states follow the same presentation format in the early and late basis.
Applying a CZ gate on the photonic states \cite{Lo20} written in the four rotated Bell states would result in 
\begin{equation}
\text{CZ}_p\ket{\psi}=(|++\rangle_{p} |\tilde{\phi}^{+}\rangle_{L} -|--\rangle_{p} |\tilde{\phi}^{-}\rangle_{L} -|-+\rangle_{p} |\tilde{\psi}^{+}\rangle_{L} - |+-\rangle_{p} |\tilde{\psi}^{-}\rangle_{L} )/2.
\end{equation}
Subsequent projective measurement of the  photonic pair in the Bell basis \cite{Val14} guarantees the entanglement of logical qubits.

 \section*{Appendix~D: Preserving the  ground motional state}
 \label{Sec_Adiabaticity}

Over the gate operation, the Rydberg-Fermi potential modifies the optical trapping experienced by the plaquette atoms in $\ket{1_t}$ Wannier states. To preserve the ground motional state, it is important to apply the $\Omega_s$ adiabatically.
Over the  operation,  plaquette atoms in  $\ket{1_t}$ Wannier state would experience trap evolution $U_{trap}=U_{op}+P_{r_c}(t)V_{RF}$ where $U_{op}$ is the optical trapp, $P_{r_c}=(\frac{\Omega_s}{\Omega_r})^2$ is the Rydberg population of the central atom.
As long as the dynamic is adiabatic, i.e. $\dot{\omega}_{trap}\ll \omega^2_{trap}$ \cite{Wit20}, the Wannier states of the $\ket{1_t}$  can  adapt continuously and stays close to the instantaneous ground motional state. For the setup discussed in the main text, the $\Omega_s$ pulses as short as 0.5$\mu$s could be designed to preserve the adiabaticity. This is significantly shorter than the Gaussian pulse with $\sigma=3.5\mu$s considered in the main text.  
Also, note that applying the dark-state optical-lattice  with better site confinement \cite{Yav09}, allows faster adiabatic operations.

%======

 \section*{S3: Closing the Rydberg-molecule decoherence channel}
In the Bose-Einstein condensate (BEC), the Fermi scattering of Rydberg electron from a free ground state atom could enhance the decoherence rate \cite{Bal13}.
This decoherence is due to attractive Rydberg-Fermi potential close to the ionic core, which moves the two interacting atoms to a very small separation of about  2 nm, where the binding energy of the molecules can ionize the Rydberg electron and form a $Cs_2^+$ molecule \cite{Nie15}. 
Confining atoms in the lattice, the interatomic separation in this scheme is tuned to be at the last lobe of Ryd-Fermi interaction, preserving the interatomic distance.
Without the mass transport, step-wise decay or ionization of the Rydberg atom is ruled out. This is because the Rydberg-Fermi binding energy at 170nm lattice constant used in this paper is orders of magnitude smaller than the closest Rydberg levels for the principal numbers applied here. The ion-pair formation is also highly unlikely in this system \cite{Nie15}.

 \section{A.~3: Addressing individual sites}

Considering the small lattice constant, single site addressing would be challenging. There are two approaches to avoid laser cross talk.  In the first approach, the sub-wavelength localized population rotation via semi interferometer techniques \cite{Aga06,Cho07}  could be used to improve single site addressing.  

    \begin{figure}[h]
\centering         \scalebox{0.48}{\includegraphics{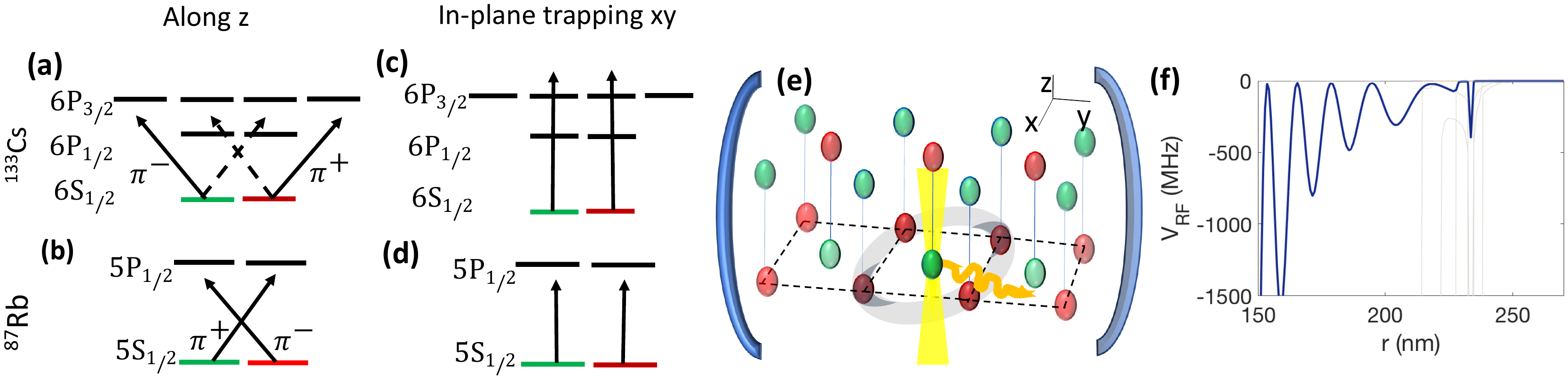}} 
\caption{ Dual-species spin-dependent lattice. (a)  The 870nm trapping laser perpendicular to the lattice plane is aligned between  $6P_{3/2}$ and $6P_{1/2}$ for Cs, hence the polarizability of qubit states $|0\rangle$ and $|1\rangle$ are given by distinguished circularly polarized lights $\pi^{-}$ and $\pi^{+}$ respectively. (b) the same polarization elements traps opposite qubit states in Rb. (c,d) For in-plane trapping the 2D standing wave is formed by 820nm linearly polarized light trapping both Cs and Rb  at the nodes and anti-nodes of the standing wave via blue and red detuned transitions.  The resulting qubit-dependent arrangement is depicted in (e) where the green and red ovals present the position of $|0\rangle$ and $|1\rangle$ qubit states respectively. 
(f) PEC with S- and P-wave scattering of the Rb atom's Rydberg electron  from a ground state Cs atom. Here the coupling of the Rydberg state $|48D_{5/2},5/2\rangle$ with the neighbouring states  $|46H+48D_{3/2}+49P_{\{1/2,3/2\}}+50S_{1/2}\rangle$ is considered under Eq.~\ref{Eq_RydFermi}. Interaction strength is plotted across radial direction with $\theta=\pi/2$ being perpendicular to the lattice plane. 
}\label{Fig_DualSpecies}
\end{figure}

In the second approach, a dual species optical lattice of $^{133}$Cs and $^{87}$Rb is considered \cite{Sin22,She21,Sin21,LHZ15}. The trapping in the lattice plane is formed by 820nm linearly polarized light that is blue and red detuned for $^{133}$Cs and $^{87}$Rb and hence they would be stored on the nodes and antinodes of the standing wave respectively, see Fig.~\ref{Fig_DualSpecies}c,d. The  qubit-dependent trapping perpendicular to the lattice plane is formed by a 870nm standing waves of left and right circularly polarized lights as depicted in Fig.~2b of the main text. The $\ket{0}$ qubit state of Rb (Cs) would be trapped on the antinodes of $\pi^{-}$ ($\pi^{+}$) standing wave and $\ket{1}$ gets trapped on the opposite polarization lattice, see Fig.~\ref{Fig_DualSpecies}a,b,e. The physical qubits are encoded in the Cs atoms placed on the plaquette while the central auxiliary atoms are Rb. This arrangement optimizes the interaction since the electron scattering from Cs atoms features resonance at lower kinetic energies close to last lobe \cite{Boo15}. The corresponding Rydberg-Fermi interaction over a single Cs site would be $\bra{0}V_{RF}\ket{0}/2\pi=50$MHz. Considering the adjusted $\Omega_s=$428nm laser for the new Rb central atom, focusing the laser beam to a single site with NA=0.8 microscope \cite{Bak09}, the probability of photon emission from the neighboring Rb atoms after an $\Omega_s$ pulse would be 1\%. The Rydberg-Fermi interaction PEC of $|48D_{5/2},5/2\rangle$ Rydberg state targeting the new inter-atomic separation of 205nm is plotted in Fig.~\ref{Fig_DualSpecies}f.

\end{widetext}


\begin{thebibliography}{100}
\bibitem{Microwave}The emitted photons could be in microwave or optical regime appropriate for both fibre and free-space communication.
\bibitem{qubit}The logical basis states for the six-qubit code  \cite{Sha08} are: \\
$\ket{0}_L =\ket{000000}-\ket{100111}+\ket{001111}-\ket{101000}-\ket{010100}+\ket{110011}+ \ket{011011} - \ket{111100}$,\\
$ \ket{1}_L=\ket{001100} + \ket{101011} + \ket{000011} + \ket{100100}-\ket{011000} - \ket{111111} + \ket{010111} + \ket{110000} $\\
which are the odd and even parity of logical operator $\bar{Z}=I Z I Z I Z$.
\end{thebibliography}
\end{document}